\documentclass[12pt,english,floatfix,superscriptaddress,aps,preprint,showpacs]{revtex4}
\usepackage{amsmath}
\usepackage{amssymb}
\usepackage{amsbsy}
\usepackage{amsfonts}
\usepackage{amsopn}
\usepackage{amstext}
\usepackage{graphicx}
\usepackage{amssymb}
\usepackage{amsfonts}
\usepackage{amsmath}
\usepackage{graphicx}
\usepackage[english]{babel}
\usepackage{color}
\usepackage[dvips]{epsfig}
\usepackage[dvips]{graphicx}
\usepackage{float}
\usepackage{units}
\usepackage{textcomp}
\usepackage{babel}

\begin{document}

\title{Kalb-Ramond field localization on the Bloch brane}
\author{W. T. Cruz}
\email{wilamicruz@gmail.com}
\affiliation{Instituto Federal de Educa\c{c}\~ao, Ci\^encia e Tecnologia do Cear\'a (IFCE),
Campus Juazeiro do Norte, 63040-000 Juazeiro do Norte - CE - Brazil }
\author{R. V. Maluf}
\email{r.v.maluf@fisica.ufc.br}
\affiliation{Universidade Federal do Cear\'a (UFC), Departamento de F\'isica, Campus do
Pici, Fortaleza - CE, C.P. 6030, 60455-760 - Brazil}
\author{C. A. S. Almeida}
\email{carlos@fisica.ufc.br}
\affiliation{Universidade Federal do Cear\'a (UFC), Departamento de F\'isica, Campus do
Pici, Fortaleza - CE, C.P. 6030, 60455-760 - Brazil}
\date{\today}

\begin{abstract}
This work deals with new results on the Kalb-Ramond (KR) field localization in braneworld models.
 We consider a five-dimensional warped spacetime with an embedded 4D thick brane which is generated  by  two real scalar fields coupled with gravity (the so called Bloch brane). We find a KR field zero mode localized with the inclusion of the dilaton coupling. Analyzing the massive spectrum, we detected a series of resonant modes  that arise from the solutions of the Schr\" odinger-like equation for KR field. The effects of the brane thickness and of the dilaton coupling over the resonance structures are determined.  Such analysis is extended to the resonance lifetimes of the massive modes, allowing a better understanding on the localization mechanism of the model.
\end{abstract}


\maketitle

\section{Introduction\label{intro}}

Over the past years, brane-world models have been proposed as an alternative solution to solve the hierarchy and the cosmological constant problems \cite{RS,add,add2}. Such models consider the gravity free to propagate into the extra dimension, while the matter fields are constrained to propagate only on the brane. However, some problems were encountered to localize the standard model (SM) fields in Randall-Sundrum (RS) scenario, such as spin-1 fields  \cite{davo}. In this way, to circumvent these problems and to keep the idea of brane worlds, some works have considered a five-dimensional spacetime (bulk) where the observable universe (brane) is taken as a topological defect with solutions that may be interpreted as non-singular versions of the RS scenario \cite{gremm,de}.

Topological defects like domain walls have been used to represent brane-world models with one extra dimension \cite{gremm,rub,vis}. Such models can be constructed from scalar fields coupled to gravity, giving rise to thick branes scenarios \cite{hoff}. Some of these models may be used to mimic new brane worlds containing internal structures
\cite{aplications} with implications on the
density of matter-energy along the extra dimensions \cite{brane}. The  advantages of these models over the RS model is that they are non-singular and  dynamically generated. Moreover, the warp factor is a smooth function determined by the scalar potential. In this work we consider a Bloch brane model \cite{dionisio}, which is constructed from two real scalar fields coupled with gravity. Such model admits internal structure and the defect thickness is controlled by a real parameter. Recently, the localization of several types of bulk fields, such as fermions \cite{castro,ca,chineses0}, gauge fields \cite{nosso6} and gravity \cite{dionisio} have been investigated on the above scenario. In these papers, however, the issue concerning the localization of the antisymmetric tensor fields, such as the Kalb-Ramond field was not regarded.

The Kalb-Ramond field is a 2-form field which plays important role in string theories and field theories. For example, it appears as a mode of massless excitations in the low-energy limit of closed strings. In extended supergravity it emerges as an auxiliary field in a supermultiplet \cite{green}. Also in string theory, the rank-2 skew-symmetric KR field is the gauge of strings, that is coupled, in heterotic string theory, to the Yang-Mills-Chern-Simons three-form \cite{green,string}.

The anti-symmetrical tensors are natural objects pertaining in differential manifolds. They give important features in the construction of the manifold's volume and therefore its orientation \cite{nakahara}.  Recently, in RS higher-dimensional framework, a lot of attention has been given to the study of optical activity of electromagnetic waves in a KR cosmological background \cite{optical}. The field strength tensor for KR field $H_{MNQ} $ is usually related with a Riemann-Cartan spacetime with torsion \cite{torsion1}. Recent studies at the RS scenario have associated the presence of the torsion with the KR field, which can exist together with gravity even outside of the brane \cite{torsion2}.  We can still cite the importance of higher spin fields in $AdS_5$ spaces due to the AdS/CFT correspondence \cite{spin}.

The existence of a tensor gauge field on the Bloch brane are of great interest because it can indicate new types of particles. It is therefore, quite important to study the dynamics of such field and, especially, their localization in the brane-world scenarios. We consider specifically the Bloch brane with the inclusion of the dilaton field because the KR field could not be localized on the basic two-field model as presented in the work \cite{dionisio}.

From previous results obtained in thick brane models, we see that the dilaton scalar field is required to be added to the background and to interact directly with gauge fields in order to obtain normalized zero mode solutions \cite{kehagias}. The same coupling was introduced in Refs. \cite{nosso,nosso1,nosso2} to study the behavior of vector and tensor gauge fields in branes generated by kinks. In string theory, the dilaton  is considered a scalar partner of the spin-2 graviton and determines the string coupling constant \cite{veneziano}. Furthermore, the dilaton-matter coupling violates the equivalence principle, as shown in Ref. \cite{veneziano2}.

In Ref. \cite{decarlos} the authors consider the dilaton and moduli supermultiplets as candidates for dark matter, showing evidences of low-energy physics in string theory. Motivated by these results, we propose to apply the same technique to analyze the behavior of the KR field on the Bloch brane. As we will see, the thickness defect will influence the coupling of the massive modes with matter, as well as the intensity of the dilaton coupling with the KR field.

Recently, a lot of attention has been given to the search of resonances in warped space-times \cite{gremm,ca,nosso6,nosso,chineses1,chineses2,chineses3,chineses4,chineses5,csaki2,csaki1}. The interest in such objects is that they give us important information about the interaction of Kaluza-Klein massive spectrum with the four-dimensional brane. In thick brane scenarios, the resulting potential on the Schr\"{o}edinger-like equation has a volcano form. So, a natural question is if the incident plane wave of the extra dimension may exhibit a resonant structure with this potential. The physical importance of these states is that to a four-dimensional observer they will give  a quasi-discrete spectrum of low mass KK states with non-suppressed couplings to the matter on the defect \cite{gremm}.

In the context of gravity localization, the existence of a resonant mode at zero energy was related to ``quasilocalization" of the graviton \cite{csaki1}. More recently, resonant modes were obtained  for fermionic fields in various thick brane models \cite{ca,chineses1,chineses2,chineses3,chineses4}. Such states also helped the analysis of the KK modes of gauge fields \cite{nosso6,nosso}. Now, we have applied the same approach to study the behavior of the KR field localization on a brane with internal structure. The resonances found reveal us a quasi-discrete spectrum of massive modes highly coupled to the brane.

This work is organized as follows. In Sec. \ref{sec.bloch}, we present the Bloch brane scenario added to the dilaton coupling. In Sec. \ref{sec.zero} we search for a zero mode localized for the KR field on the previous setup. In Sec. \ref{sec.res}, we analyze the massive spectrum and search for resonance structures. Finally, we present our conclusions in Sec. \ref{sec.conc}.

\section{Dilatonic Bloch brane setup\label{sec.bloch}}

In this section we introduce the thick brane scenario that we consider to study the localization of tensor gauge fields. The basic Bloch brane background, as previously studied in references \cite{dionisio,castro,ca},  is unable to localize the KR field zero mode \cite{nosso1}. Indeed, the tensor gauge field geometry associated with the warp factor leads to a divergent effective action. Thus, an interesting question is the investigation of how the KR field can be localized. In this context, a mechanism largely used to allow the localization of fields in thick brane setups \cite{kehagias,nosso,nosso1,nosso2,mk3} consists in adding  the dilaton field in the matter sector. So, we will study the features of the KR field in a scenario  named {\it dilatonic Bloch brane},  defined by the action
\begin{eqnarray}\label{acdil}
\mathcal{S}=\int d^5x\sqrt{-G}\bigg[\frac{1}{4}R-\frac{1}{2}(\partial\phi)^2-\frac{1}{2}(\partial\chi)^2-\frac{1}{2}(\partial\pi)^2-V(\phi,\chi,\pi)\bigg],
\end{eqnarray}
where $G=\det (G_{AB})$ and $R$ is the Ricci scalar, with Latin indices  used for the bulk coordinates. The brane is composed by the two fields $\phi$ and $\chi$ which depend only on the extra dimension $y$.  Following reference \cite{kehagias}, we maintain the fields $\phi$ and $\chi$  and include the dilaton scalar field $\pi(y)$ that will couple directly with the KR field. It is important to note that the scalar fields form the background, however, they are not really bulk fields but the stuff of which the brane is made. For this reason we will not analyze the localization of the scalar fields like the dilaton. Instead, we focus our attention on the localization of the bulk field considered, the tensor gauge field. The differences of the dilaton to the other scalar fields will be clarified when we complete the background description.  The geometric framework is an AdS five-dimensional spacetime with the following metric:
\begin{equation}\label{metric}
ds^2=e^{2A(y)}\eta_{\mu\nu}dx^{\mu}dx^{\nu}+e^{2B(y)}dy^2,
\end{equation} where $A(y)$ and $B(y)$ are warp factors to be determined by the Einstein equations.  The Minkowski  metric $n_{\mu\nu}$ is diagonal with entries $(-1,1,1,1)$ and Greek indices vary from $1$ to $4$.
The equations of motion coming from the action are
\begin{eqnarray}\label{eqmov}
\phi'^{2}+\chi'^{2}+\pi'^{2}-2e^{2B}V&=&6A'^{2}\nonumber\\
\phi'^{2}+\chi'^{2}+\pi'^{2}+2e^{2B}V&=&-6A'^{2}+3A'B'-3A''\\
\gamma''+(4A'-B')\gamma'&=& e^{2B}\partial _{\gamma}V, \ \mbox{with} \ \gamma=\phi,\chi,\pi,\nonumber
\end{eqnarray}
where prime stands for derivative with respect to $y$.

Therefore, we can obtain first-order differential equations from  Eq. (\ref{eqmov}) by taking the potential in terms of a superpotential $W(\phi,\chi)$ \cite{bazeia1,bazeia2,bazeia3}. To maintain the Bloch brane structure we consider the following potential function \cite{kehagias}:
\begin{equation}
V=e^{\pi \sqrt{\frac{2}{3}}}\left[\frac{1}{8}\left(\frac{\partial W}{\partial \phi}\right)^2+\frac{1}{8}\left(\frac{\partial W}{\partial \chi}\right)^2 -\frac{5}{16}W^2\right].
\end{equation}
We obtain first-order differential equations that solve the equations of motion
\begin{eqnarray}\label{singularidade1}
\phi^{\prime}&=&\frac12\,\frac{\partial W}{\partial\phi}, \ \ \
\chi^{\prime}=\frac12\,\frac{\partial W}{\partial\chi}\nonumber\\
\pi&=&-\sqrt{\frac{3}{8}}A,\nonumber\\
B&=&\frac{A}{4}=-\frac{\pi}{2}\sqrt{\frac{2}{3}},\nonumber\\
A'&=&-\frac{W}{3}.
\end{eqnarray}
The Bloch brane model is obtained by the following superpotential function \cite{bazeia4,shif,alonso}:
\begin{equation}\label{w}
W(\phi,\chi)=2\phi-\frac23\phi^3-2r\phi\chi^2,
\end{equation}
and we obtain the solutions to the scalar fields modeling the brane as
\begin{equation}\label{phi}
\phi(y)=\tanh(2ry),
\end{equation}
and
 \begin{equation}\label{chi}
\chi(y)=\sqrt{\left(\frac1{r}-2\right)}\;{\rm sech}(2ry).
\end{equation}
The factor $r$ is a real parameter that controls the brane thickness. We can see that for the limit $r=\frac{1}{2}$, the one-field scenario is recovered.

The warp factor obtained is
\begin{equation}\label{warp}
A(y)=\frac{1}{9r}\Bigl[(1-3r)\tanh^2(2ry)-2\ln\cosh(2ry)\Bigr].
\end{equation}
This brane model supports internal structure which influences the matter-energy density along the extra dimension. We can find more details about this feature and graphical analysis of the solutions (\ref{phi}, \ref{chi}, and \ref{warp}) in the reference \cite{dionisio}.

Despite the inclusion of the dilaton scalar field, our brane model preserves the same solutions as the original Bloch brane scenario. The complete analysis showing the stability of this thick brane model was performed in the work \cite{dionisio}. Furthermore, the same soliton solution was also obtained in the study of systems of two coupled scalar fields \cite{stable1,stable2}, and have motivated the construction of the Bloch brane.

The study concerning the classical stability of solutions obtained from systems of coupled scalar fields, as we have considered, was performed in the works \cite{stable3,stable4}. In such approaches the classical or linear stability was shown by proving that the associate Schr\"{o}dinger operator is positive semi-definite.  After performing fluctuations about static solutions of the scalar fields and substituting in the Euler-Lagrange equations of motion, Bazeia and collaborators obtained a Schr\"{o}dinger equation with non-negative eigenvalues.

The metric that we use with the dilaton coupling, namely the metric defined by Eq. (\ref{metric}), is different from that it is used in the original Bloch brane setup. However, as showed in Ref. \cite{kehagias}, the standard perturbation in that metric leads to graviton equations that can be cast into a supersymmetric quantum mechanical form ensuring us that their spectrum is free from tachyons.

\section{Kalb-Ramond field zero mode\label{sec.zero}}

\begin{figure}
\resizebox{0.67\textwidth}{!}{
  \includegraphics{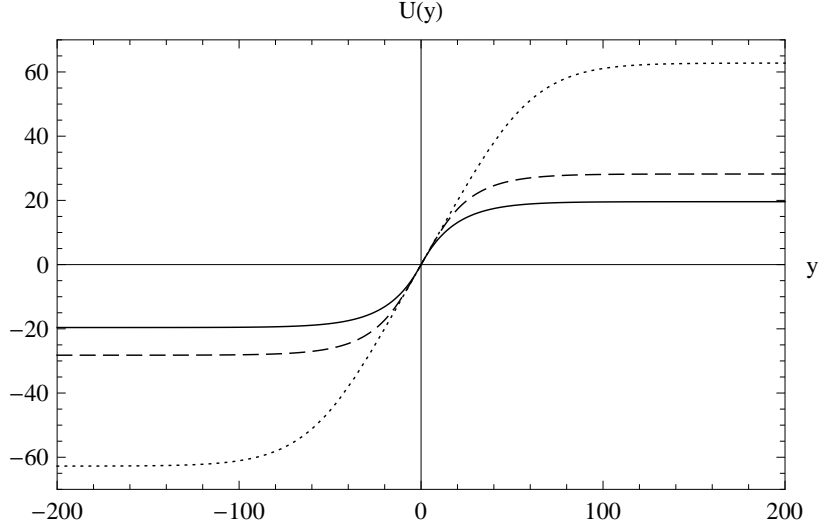}
}
%\vspace{5cm}       % Give the correct figure height in cm
\caption{Plots of the solution $U(y)$ in Eq. (\ref{geral}) with $r=0.01$ (points), $r=0.05$ (dashed line) and $r=0.25$ (solid line). We have put $\lambda=1/8$.}
\label{fig.1}       % Give a unique label
\end{figure}

After define the brane scenario, the next step is to analyze the localization of the KR field zero mode and the influence of the dilaton coupling. The dilaton field will couple directly with the KR field in a combination defined by the action \cite{dilaton1,dilaton2}:
\begin{equation}
S\sim\int d^{5}x\sqrt{-G}(e^{-\lambda\pi\sqrt{\frac{8}{3}}}H_{MNL}H^{MNL}),
\end{equation}
where $H_{MNL}=\partial_{[M}B_{NL]} $ and $\lambda$ is a constant that controls the intensity of the dilaton coupling.
Therefore, we must analyze the equations of motion for  the tensor
gauge field in the dilatonic Bloch brane background. This equation takes the form
\begin{equation}
\partial_{M}(\sqrt{-G}G^{MP}G^{NQ}G^{LR}e^{-\lambda\pi\sqrt{\frac{8}{3}}}H_{PQR})=0.
\end{equation}
With the gauge choice $B_{\alpha 5}=\partial\mu B^{\mu\nu}=0$ and
with the separation of variables
\begin{equation}
B^{\mu\nu}(x^{\alpha},y)=b^{\mu\nu}(x^{\alpha})U(y)=b^{
\mu\nu}(0)e^{ip_{\alpha}x^{\alpha}}U(y),
\end{equation}
where $p^{2}=-m^{2}$, it is obtained a differential equation which give us information about the
extra dimension, namely
\begin{eqnarray}\label{zero}
\frac{d^{2}U(y)}{dy^{2}}-\big[\lambda\pi^{\prime}(y)\sqrt{\frac{8}{3}}+B^{\prime}(y)\big]\frac{dU(y)}{dy}=-m^{2}e^{2[B(y)-A(y)]}U(y)
%\label{U1}.
\end{eqnarray}
To solve Eq. (\ref{zero}) for zero mode $m=0$, we use the following relation
\begin{equation}
\frac{dU(y)}{dy}=g(y),
\end{equation}
and obtain a solution for $g(y)$ as
\begin{equation}
g(y)=k e^{[\lambda \pi(y)\sqrt{\frac{8}{3}}+B(y)]}, \,\,\, k=g(0)e^{-[\lambda \pi(0)\sqrt{\frac{8}{3}}+B(0)]}.
\end{equation}
Using the relations (\ref{singularidade1}) we find a solution for the zero mode of the KR field as a function of the extra dimension given by
\begin{equation}\label{geral}
U(y)=k\int^{y}_{y_0}e^{(\frac{1}{4}-\lambda)A(y')}dy'.
\end{equation}
The above solution is finite for $\lambda<\frac{1}{4}$ where it interpolates between two constant values on both sides of the membrane when $y\rightarrow\pm\infty$. The solution (\ref{geral}) is plotted in Figure (\ref{fig.1}). As we can see, there is a transition region which separates the two interfaces. The thickness of this region as well as the thickness of the defect is controlled by the parameter $r$. In this case, when increasing $r$ the thickness of the transition region is reduced.

Despite we obtain a finite solution, we still cannot guarantee the existence of a localized zero mode.
The effective action for the zero mode in five dimensions is
\begin{eqnarray}
S\sim\int d^{5}x\sqrt{-G}(e^{-\lambda\pi\sqrt{\frac{8}{3}}}H_{MNL}H^{MNL})=\int dy
(U(y))^{2}e^{A(y)(\lambda-\frac{7}{4})}\int d^{4}x(h_{\mu\nu\alpha}h^{\mu\nu\alpha}).
\end{eqnarray}
We observe that for $\lambda<\frac{1}{4}$, the condition to have a finite solution for $U(y)$, we obtain a divergent effective action.
However, a particular solution of the equation of motion (\ref{zero})
is simply $U(y)\equiv \mbox{cte}$. Given the solution $U(y)$ constant it is possible to show that the
integral in the $y$ variable on the effective action is finite if $\lambda >
\frac{7}{4}$. As a consequence, for
a specific value of the coupling constant $\lambda$ it is possible
to obtain a localized zero mode of the KR field.

In order to show the importance of choosing the dilatonic Bloch brane to have a localized zero mode we can test cancel the dilaton influence. To this purpose we consider $\lambda=0$ and the metric $ds^{2}=e^{2A(y)}\eta_{\mu\nu}dx^{\mu}dx^{\nu}+dy^{2}$. With the same gauge choice and separation of variables, we find the following equation to the KR field in the extra dimension:
\begin{equation}\label{motion}
\frac{d^2U(y)}{dy^2}=-m^2e^{-2A(y)}U(y)
\end{equation}
When $m^2=0$ we have the solutions $U(y)=cy + d$ and $U(y)=c$ with
$c$ and $d$ constants. We take again the effective
action for the tensor field where we decomposed the part dependent on
the extra dimension,
\begin{eqnarray}
S\sim\int dy U(y)^2
e^{-2A(y)} \int d^4 x(h_{\mu\nu\lambda}h^{\mu\nu\lambda}).
\end{eqnarray}
Given the solutions for $A(y)$ in Eq. (\ref{warp}) and for $U(y)$ obtained above, we
clearly observe that due to the minus sign in the warp factor, the
function $U(y)^2 e^{-2A(y)}$ goes to infinity for the two
solutions of $U(y)$. In this way, the effective action for the zero
mode of the Kalb-Ramond field is not finite after integrating the
extra dimension. This show us that the Bloch brane scenario without the dilaton coupling is  not capable to support a zero mode localized to the KR field.

\section{Massive spectrum and resonances}\label{sec.res}

In order to analyze the massive modes in this
background, we make use of Eq.  (\ref{zero})  and the relations  obtained in Eq. (\ref{singularidade1}). Thus,  we get
\begin{equation}\label{massive}
\frac{d^{2}U(y)}{dy^{2}}-\alpha A'(y)\frac{dU(y)}{dy}=-m^{2}e^{-\frac{3}{2}A(y)}U(y),
%\label{U1}.
\end{equation}
where $\alpha= \frac{1}{4}-\lambda$. Performing the following transformations:
\begin{equation}\label{trans2}
dz=dy e^{-\frac{3}{4}A},\;\;\;  U=e^{\left(\frac{\alpha}{2}+\frac{3}{8}\right)A}\overline{U},
\end{equation}
we can write Eq. (\ref{massive}) as a Schr\"odinger-like equation  in the form
\begin{equation}\label{schro}
\left\{-\frac{d^2}{dz^2}+\overline{V}(z)\right\}\overline{U}=m^2\overline{U},
\end{equation}
with the potential $\overline{V}(z)$  given by
\begin{equation}\label{vp}
\overline{V}(z)=\left[\beta^2(\dot{A})^2-\beta\ddot{A}\right],\;\;\;\;
\beta=\frac{\alpha}{2}+\frac{3}{8},
\end{equation} where the dot stands for derivative with respect to $z$.

The Schr\"{o}dinger-like equation
(\ref{schro}) can be rewritten in the formalism of supersymmetric quantum mechanics as follows:
\begin{equation}\label{susy_qm}
Q^{\dag} \, Q \,
\overline{U}(z)=\left\{-\frac{d}{dz}+\beta\dot{A}\right\}\left\{\frac{d}{dz}+\beta\dot{A}\right\}\overline{U}(z)=m^2\overline{U}(z).
\end{equation}
Now looking at the form of Eq. (\ref{susy_qm}), we can exclude the possibility
that  the normalized negative energy modes exist \cite{bazeia0}.

The mass spectrum is determined by the characteristics of the potential at infinity. If $\overline{V}\rightarrow 0$ when $z\rightarrow\infty$, then we have a continuous gapless spectrum of Kaluza-Klein states. Due to the transformations in Eq. (\ref{trans2}), we cannot achieve the analytical structure of the potential $\overline{V}(z)$ and hence, the wave functions for the massive modes that are solutions of Eq. (\ref{susy_qm}) does not have an analytical representation.  However, we will be able to analyze $\overline{V}(z)$ numerically.

\begin{figure*}
\centering
\resizebox{1\textwidth}{!}{%
  \includegraphics{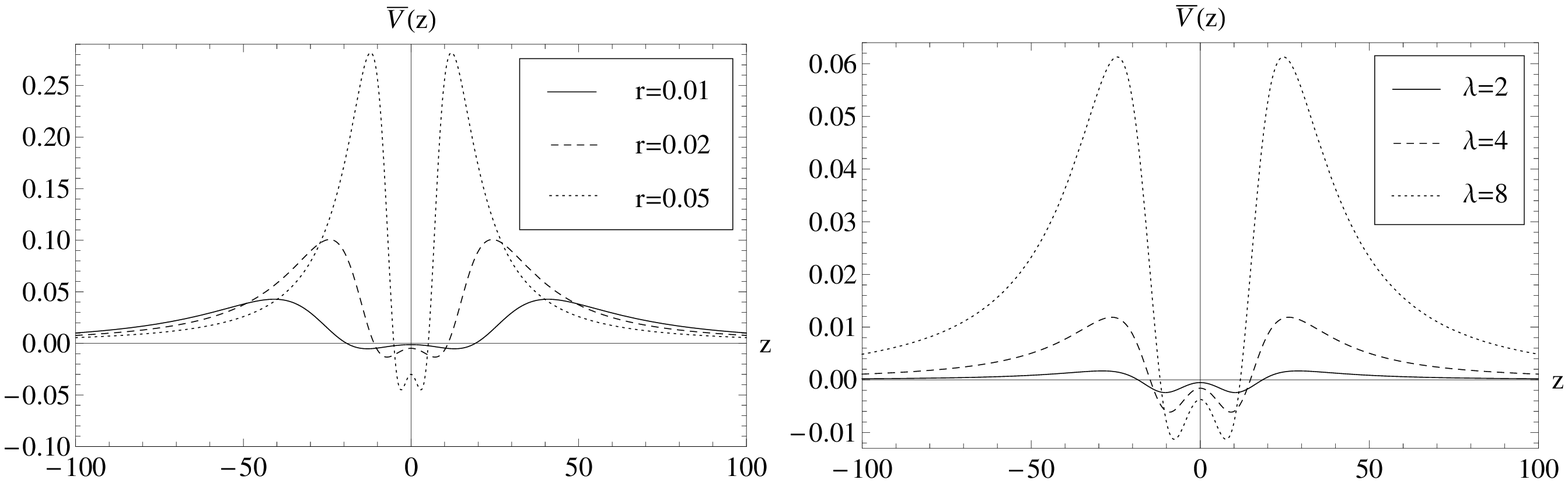}
}
\vspace{0cm}       % Give the correct figure height in cm
\caption{Plots of $\overline{V}(z)$ for $\lambda=10$ (left) and $r=0.01$ (right).}
\label{kr1}       % Give a unique label
\end{figure*}

In Fig. \ref{kr1} we have shown $\overline{V}(z)$ for some values of constants $\lambda$ and $r$. Observing the
asymptotic behavior of the potential, we conclude that $\lim_{z \rightarrow \infty}{\overline{V}(z)} =0$
and the continuum spectrum is gapless. The most relevant information that we can extract from these graphs refers to the maximum points of the potential. The solutions of massive modes on the brane depends on the maximum value of the potential, but for $m^2\gg\overline{V}_{\mbox{max}}$ this dependence becomes negligible. However, when we have $m^2<\overline{V}_{\mbox{max}}$ the potential structure will influence the solution of the massive modes near $z=0$.

Looking still to Fig. \ref{kr1}, we see that raising the value of the dilaton coupling constant implies an increasing in the maximum value of the potential. The same feature is observed when increasing the value of $r$, i.e., when one grows up the thickness of the brane.

\begin{figure*}
\centering
\resizebox{1\textwidth}{!}{%
  \includegraphics{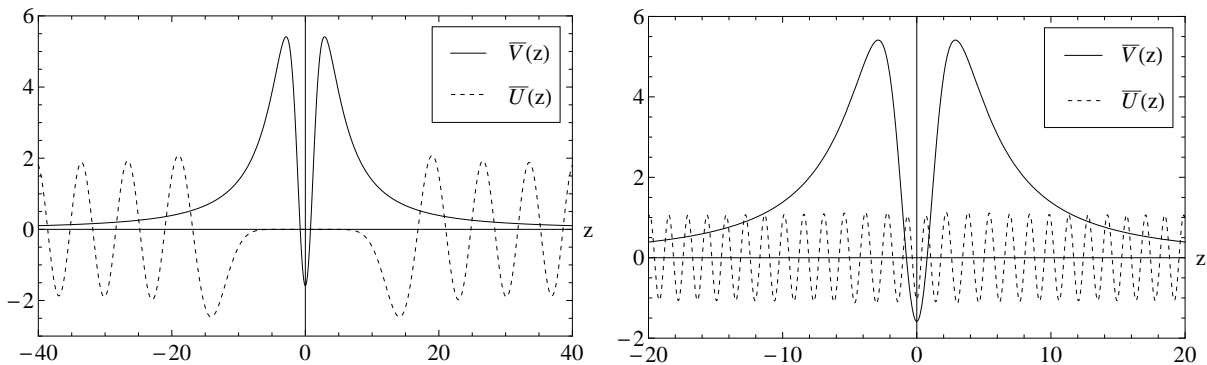}
}
\vspace{0cm}       % Give the correct figure height in cm
\caption{Plots of $\overline{V}(z)$ and $\overline{U}(z)$ for $m=1$ (left) and $m=20$ (right). We put $\lambda=20$ and $r=0.25$ in two graphics.}
\label{kr0}       % Give a unique label
\end{figure*}

For a given potential we must know the behavior of the solutions $\overline{U}(z)$ in relation to their masses. With this purpose,  we plot in Fig.  (\ref{kr0})  the  solutions of $\overline{U}(z)$ for  $m=1$ (left) and $m=20$ (right). In both cases, we use the same potential, while keeping  constant $\lambda$ and $r$. The light modes  ($m^2<\overline{V}_{\mbox{max}}$) are suppressed  in the region between the maxima of the potential and in the region distant from the brane, where they can be approximated by plane wave solutions. However, the modes in which $m^2\gg\overline{V}_ {\mbox{max}}$ possess plane wave solution across all extra dimension. It is worth noting that either the dilaton coupling as well as  the brane thickness affect the solutions of the massive modes in the extra dimension. Thus, these two constants modify the coupling of the tensor gauge field with the matter on the brane.
%\begin{figure*}
%\includegraphics[scale=0.6]{kr0.eps}
%\includegraphics[width=0.75\textwidth]{kr0.eps}
%\caption{Plots of $\overline{V}(z)$ and $\overline{U}(z)$ for $m=1$ (left) and $m=20$ (right). We put $\lambda=20$ and $r=0.25$ in two graphics.}
%\label{kr0}

A very interesting feature is that although the massive modes with $m^2<\overline{V}_{\mbox{max}}$ are suppressed at $z=0$, there is a possibility that for some specific values of masses, the plane wave solutions of the Schr\"odinger equations can resonate with the potential, yielding resonance modes \cite{gremm,csaki1,csaki2}. Such structures exhibit very large solutions within the brane, when comparing its value when $z\rightarrow\infty$.  In order to search for resonances, we must know the solutions of $\overline{U}(z)$ at $z=0$ with respect to the mass.  In this way,  we  take the probability density of the wave function at the center of the brane $P(m)=|\zeta\overline{U}(0)|^2$, where $\zeta$ is a normalization constant taken at the box $-200<z<200$.

\begin{figure}
\centering
\resizebox{0.7\textwidth}{!}{%
  \includegraphics{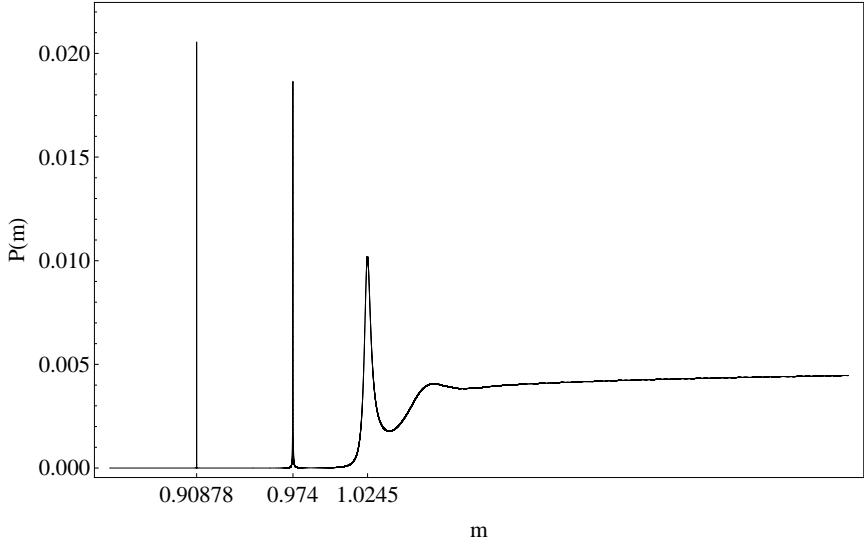}
}
\vspace{0cm}       % Give the correct figure height in cm
\caption{ Plots of $P(m)$ with $\lambda=46$ and $r=0.01$. We note three resonance peaks with masses $m=0.90878$, $m=0.9740$ and $m=1.0245$ with lifetimes $\tau=1.00\times10^{5}$, $\tau=5.55\times10^{3}$ and $\tau=1.76\times10^{2}$ respectively.}
\label{kr2}       % Give a unique label
\end{figure}

In Fig. (\ref{kr2}) we present $P(m)$ for $\lambda=46$ and $r=0.01$. It is observed the existence of three resonances in $m=0.90878$, $m=0.9740$ and $m=1.0245$. The difference among them, besides the masses, is determined by their lifetimes. We can estimate the lifetime $\tau$ of a resonance by $(\Delta m)^{-1}$, where $\Delta m$ is the width at half maximum in mass of the peaks in $P(m)$ \cite{ca,chineses1,lifetime,chineses2,chineses3}.

%\begin{figure}
%\includegraphics[scale=0.3]{kr2.eps}
%\caption{ Plots of $P(m)$ with $\lambda=46$ and $r=0.01$. We note three %resonance peaks with masses $m=0.90878$, $m=0.9740$ and $m=1.0245$ with %lifetimes $\tau=1.00\times10^{5}$, $\tau=5.55\times10^{3}$ and %$\tau=1.76\times10^{2}$ respectively.}\label{kr2}
%\end{figure}

\begin{figure*}
\centering
\resizebox{1\textwidth}{!}{%
  \includegraphics{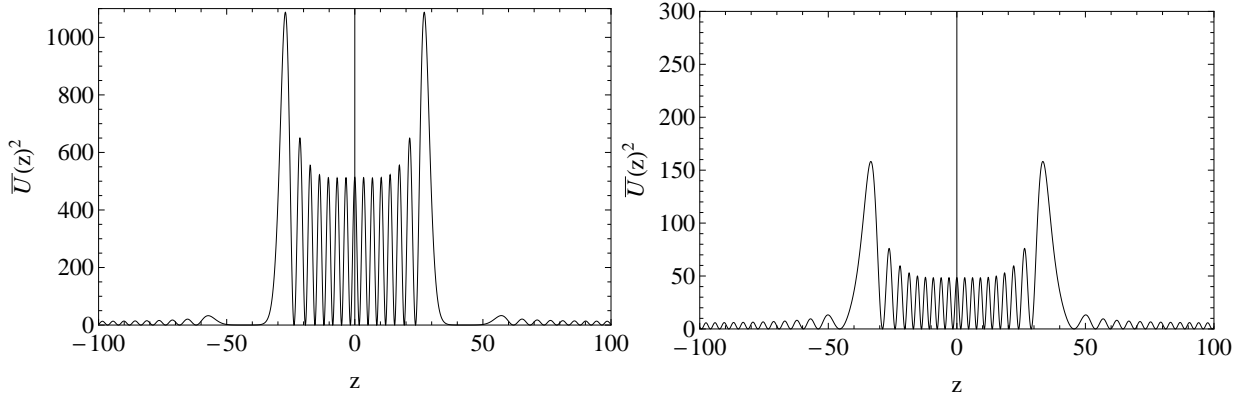}
}
\vspace{0cm}       % Give the correct figure height in cm
\caption{Plots of $\overline{U}(z)^2$ with masses $m=0.90878$ (left) and $m=1.0245$ (right) where we have put $\lambda=46$ and $r=0.01$. Note that these two masses correspond to resonances as shown in Fig. (\ref{kr2}).}
\label{kr4}       % Give a unique label
\end{figure*}

Regarding solutions of the  Schr\"{o}dinger equation, resonances with higher lifetimes characterize solutions of $\overline{U}(z)$ with large amplitudes on the brane location. To better understand this feature, we present in Fig.  (\ref{kr4}) the solutions for two resonant modes as seen in  Fig.  (\ref{kr2}). The selected masses were $m=0.90878$ and $m=1.0245$ with lifetime $\tau=1.00\times 10^{5}$ and $\tau=1.76\times 10^{2}$, respectively. The solution for the resonant mode with the highest lifetime (left) has a wider amplitude at $z=0$,  when compared with another one (right). This shows us that these modes are highly coupled with matter on the brane, as compared to the nonresonant modes. The calculation of $\tau$ is also useful to show which peaks in $P(m)$ can be considered resonances. For example, in situations where $\Delta m$ greater than the peak position $m$, the lifetimes are so small that they do not produce any measurable effects. Thus, these modes cannot be considered resonances.

%\begin{figure*}
%\includegraphics[scale=0.5]{kr4.eps}
%\caption{Plots of $\overline{U}(z)^2$ with masses $m=0.90878$ (left) and %$m=1.0245$ (right) where we have put $\lambda=46$ and $r=0.01$. Note that %these two masses correspond to resonances as shown in Fig. (\ref{kr2}).}
%\label{kr4}
%\end{figure*}

\begin{figure*}
\centering
\resizebox{1\textwidth}{!}{%
  \includegraphics{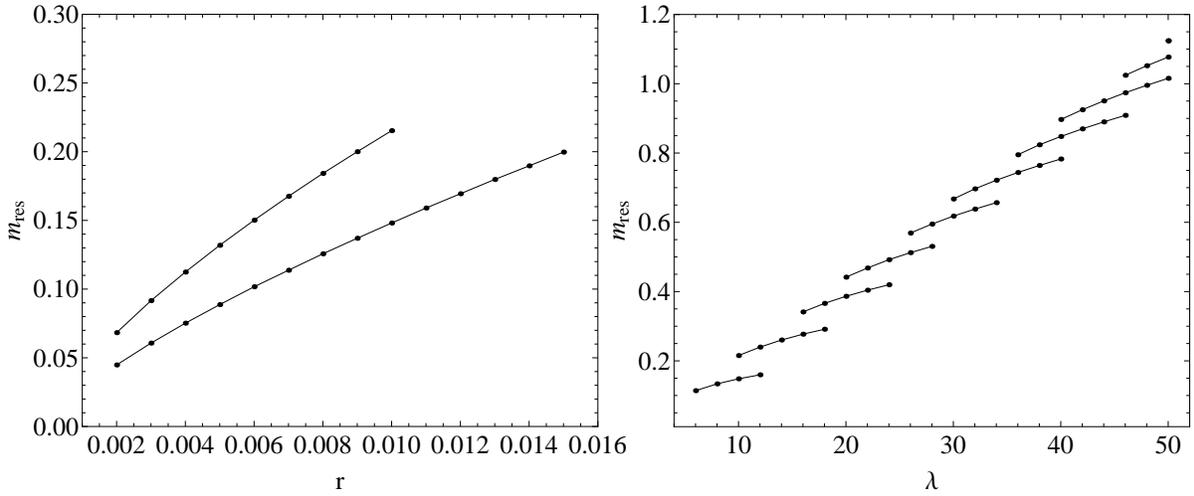}
}
\vspace{0cm}       % Give the correct figure height in cm
\caption{Plots of the resonance masses in function of $r$ (left) and $\lambda$ (right). We have put $\lambda=10$ (left) and $r=0.01$ (right).}
\label{kr3}       % Give a unique label
\end{figure*}

Another question is whether the resonance structures may be affected by the intensity of the dilaton coupling, as well as the thickness of the defect. This issue can be better understood by knowing how the masses of the resonant modes are modified according to the constants $\lambda$ and $r$. Excluding cases with very small lifetimes, we shown in Fig. (\ref{kr3}) a series of mass values of peaks obtained in terms of  $P(m)$ as a function of $\lambda$ and $r$. The connected points correspond to the same resonance structure. As we can see, increasing  the values of $\lambda$ or $r$ the masses of the peaks in the probability density function increases too.  Therefore, raising $\lambda$ or $r$ implies an increase of the mass in each resonance  spectrum.  In general, the mass range where the resonances appear correspond to larger values of $m$, as $r$ or $\lambda$ increase. This is connected to the characteristics of the potential $\overline{V}(z)$ to the KR field. The maximum potential is raised when we increase $\lambda$ or $r$ so that the light modes are suppressed in the region near $z=0$ and resonant modes appear for bigger masses.

Let us now consider what happens when the defect thickness is changed on the resonant modes. First we note that for $r=0.002$ (See Fig. \ref{kr3}, left)  there are two resonances in the spectrum. However, when $r=0.015$ there is only one resonance. The one with greater mass, represented by the upper line in Fig. \ref{kr3} (left), it vanishes near $r=0.01$. On the other hand, increasing the value of $\lambda$ as seen in Fig.\ref{kr3} (right), we observe a slight uptick in the number of resonances. This behavior can be understood through the relation on lifetimes of resonance with respect to the parameters $r$ and $\lambda$. In Table \ref{t}, we display some values of $r$ and $\lambda$, and the corresponding lifetimes. As can be seen, an increase in $r$ corresponds to a reduction on the resonance lifetimes,  and the opposite happens for $\lambda$. This feature explains why we have more resonances when increasing $\lambda$ or reducing $r$ \cite{ca}.

Still in Fig. (\ref{kr3}) we note that when we reduce $\lambda$, the lines corresponding to the same resonance disappear. This occurs because they acquire too small lifetimes. However, increasing $\lambda$, the resonance lifetimes are increased and their widths become so small that cannot be detected due to computational limitations.

\begin{table*}
\begin{ruledtabular}
\centering
\caption{The mass, width and lifetimes of resonances in function
of $\lambda$ and $r$.}
\label{t}       % Give a unique label
\begin{tabular}{cccccccc}
%\hline
\multicolumn{4}{|c||}{$\lambda=$10} & \multicolumn{4}{c|}{r=0.01}\tabularnewline
\hline

r  & m  & $\Delta m$  & $\tau$  & $\lambda$  & m  & $\Delta m$  & $\tau$ \tabularnewline
\hline
0.011  & 0.1590  & 0.00018  & 5555.55  & 30  & 0.6673  & 0.0135  & 74.074 \tabularnewline
\hline
0.012  & 0.1694  & 0.00021  & 4761.90  & 32  & 0.6966  & 0.0035  & 285.714 \tabularnewline
\hline
0.013  & 0.1798  & 0.00025  & 4000  & 34  & 0.7215  & 0.00084  & 1190.47 \tabularnewline
\hline
0.014  & 0.1897  & 0.00028  & 3571.4  & 36  & 0.7950  & 0.00017  & 5882.35 \tabularnewline
\hline
0.015  & 0.1997  & 0.00033  & 3030.3  & 38  & 0.8237  & 0.000027  & 37037.4 \tabularnewline
%\hline
\end{tabular}
% Or use
\vspace*{0cm}  % with the correct table height
\end{ruledtabular}
\end{table*}

\section{Summary}\label{sec.conc}
Concerning the localization in thick brane models,  many studies show that the KR field cannot have zero mode localized on the usual Bloch brane setup \cite{dionisio,ca,castro}. Actually, in these papers it was found that  the solutions of the equation of motion for the KR field yield a divergent effective action, breaking the localization mechanism.

Aiming to overcome this result, we employ a well-known method within the context of the localization for gauge and tensor gauge fields on the branes \cite{kehagias,nosso,nosso1,mk3}. It consists of adding an extra field to the model, the dilaton, which coupled to the KR field, enabled us to locate a zero mode on the Block brane.

Moreover, we rewrite the equation of motion for the KR field as a Schr\"odinger-like equation, and from the analysis of the massive modes we conclude that the spectrum is free of negative energy states. In particular, light mode states are suppressed in the region among the maxima of the potential. Nevertheless, there are some solutions for certain masses values that have large amplitudes at $z=0$, and which can be characterized as being resonances.

To detect the resonant modes we use a technique widely applied in the context of thick branes \cite{ca,nosso6,nosso1,nosso2,chineses1,chineses2,chineses3,chineses4,chineses5,nosso3,nosso4,nosso5} based on the calculation of the probability density of the wave function. We have normalized the wave functions in a box so that its borders are far from the turning points. As noted in Ref. \cite{ca}, the choice of the size of the box has no physical significance, since if it is chosen sufficiently  large, it does not interferes with the position of the resonant modes.

To complete the study of the massive spectrum we calculate the value of the lifetimes of resonant modes in addition to analyzing the conditions under which they can be changed. As noted, the lifetime value helps us to know the shape of the resonant modes solution as well as its coupling with the membrane. Narrow peaks on the probability density function correspond to higher lifetimes, so that they characterizes large amplitudes at $z=0$ for the solutions of the Schr\"{o}edinger equation.

We find interesting relations between the resonance life-times and the internal structure of the defect. This characteristic was also observed for the resonances found in the study of fermions localization in Bloch branes \cite{ca}. When we make the brane thicker, the resonance lifetime is reduced and it is washed out from the spectrum. Therefore, its effect will not be important. However, when we increase the coupling of the dilaton with the KR field, the width of the resonances fall indicating larger lifetimes and hence, modes highly coupled to the brane. With respect of the specific energies where the resonant modes appear, they are increased while raising the thickness of the defect or the dilaton coupling.

\begin{acknowledgments}
The authors thank the Funda\c{c}\~{a}o Cearense de apoio ao Desenvolvimento
Cient\'{\i}fico e Tecnol\'{o}gico (FUNCAP), the Coordena\c{c}\~{a}o de Aperfei\c{c}oamento de Pessoal de N\' ivel Superior (CAPES), and the Conselho Nacional de Desenvolvimento Cient\' ifico e Tecnol\' ogico (CNPq) for financial support.
\end{acknowledgments}

\end{document}